\documentstyle[openbib,12pt,fleqn]{article}
\newcommand{\bea}{\begin{eqnarray}}
\newcommand{\beq}{\begin{equation}}
\newcommand{\eea}{\end{eqnarray}}
\newcommand{\eeq}{\end{equation}}

\renewcommand{\baselinestretch}{2}

\begin{document}
\bibliographystyle{unsrt}

\setcounter{footnote}{0}

%%%%%%%%%%%%%%%%%%%%%%%%%%%%%%%%%%%%%%%%%%%%%%%%%%%%
%\tolerance = 10000
%\documentstyle[preprint,revtex]{aps}
%\baselineskip.4cm
%\hoffset -0.9truecm
%
% FINE MACRO GLOBALI --- INIZIO MACRO LOCALI
%
%\def\bref#1{[\cite{#1}]}
%\def\riga#1{\par\hfill\break{#1}\hfill\break\par}
%\begin{document}
%\draft
%
%         TITLE PAGE
%
%\pagestyle{empty}
\begin{center}
\phantom{.}
%\vskip 5.5truecm
{\Large \bf 
%%BeginAbstract
Kramers map approach for stabilization of hydrogen atom  
%%EndAbstract
\\}
{\Large \bf 
%%BeginAbstract
in a monochromatic field  
%%EndAbstract
\\}
%\vspace{-0.15truecm}
{\small \sl   
%%BeginAbstract
D.L.SHEPELYANSKY
~$^{(a)}$ 
%%EndAbstract
 \\}
%\vspace{-0.15truecm}
%\vspace{-0.15truecm}
%{\small \sf DYSCO--007}\\
%{\small PREPRINT \sf FNT/T 92/35 PV}\\
%\vspace{1.0truecm}
{\small \it 
%%BeginAbstract
Laboratoire de Physique Quantique,
  Universit\'e Paul Sabatier, \\
  118, route de Narbonne, 31062 Toulouse Cedex, France
%%EndAbstract
  }\\

\vspace{0.5truecm}

\vskip .3 truecm

\vspace{0.5truecm}
\end{center}
\small
{\bf Abstract:\/}
{
%%BeginAbstract
The phenomenon of stabilization of highly excited states of hydrogen atom
in a strong monochromatic field is discussed. Approximate description of 
dynamics by the introduced Kramers map allows to understand the main 
properties of this phenomenon on the basis of analogy with the Kepler
map. Analogy between the stabilization and the channeling of particles 
in a crystal is also discussed.
%%EndAbstract
}
\vskip .4 truecm
\vspace{0.5truecm}
{Submitted to Physica D, February/March 1993}\\

\newpage

%%%%%%%%%%%%%%%%%%%%%%%%%%%%%%%%%%%%%%%%%%

%\noindent{June 1992}
\newpage
\section{Introduction}
\indent

During the last years the phenomenon
of stabilization of atom in a strong laser field attracted a great deal
of attention \cite{ALL}. While the existence of the stabilization of 
atom has been clearly demonstrated in the numerical experiments
the clear analytical criterion of stabilization is still absent.
Usually it is assumed that stabilization 
condition is satisfied if the energy of the laser photon is larger
than the electron coupling energy and the amplitude of electron oscillations
in the field is large in comparison with Bohr radius \cite{GAV}.
However, the recent investigations of the corresponding classical
problem demonstrated that stabilization remains also in the classical
atom \cite{DS,BCS}, where the above conditions are violated.
The physical explanation of this phenomenon and the condition
of stabilization were given in \cite{DS,BCS} but the detailed 
explanation of the
effect still remains an open problem.
For a better understanding of this stabilization I introduce
here a one-dimensional atom model which I will
call Kramers model (having in mind that it
arose from the Kramers - Henneberger transformation).
Numerical analysis of this model
allowed to construct an approximate Kramers map which describes the 
process of energy excitation and gives conditions of classical ionization.
In some sense the obtained Kramers map is quite close to the
Kepler map \cite{EEE} which describes the motion in the limit of
relatively small field. Indeed, even in the strong
field the change of the electron energy
happens only when the electron passes near the nucleus while far from it
the electron follows the Kepler orbit. 
 
\indent
The paper is constructed as follows. In the section 2 a brief 
description of the Kepler map is given since analogy with this
map can be useful in the stabilization regime. In the section 3 a qualitative
explanation of the stabilization is presented. The numerical
analysis of the introduced one-dimensional Kramers model and
the derivation of the Kramers map are carried out in the section 4. 
In the section 5 I discuss
analogy between the stabilization and the channeling of electrons 
in the crystal. In the conclusion the possibilities of experimental 
observation of stabilization of Rydberg atoms are discussed.

\section{Kepler Map}
\indent
\par
After the pioneer experiments of Bayfield and Koch in 1974 \cite{1974}
the problem of microwave ionization of highly excited states
of the hydrogen atom has been investigated by many groups
(see \cite{JEN} and Refs. therein). The fast ionization observed 
in the experiments was really surprising since about 100 photons were
required to ionize the atom. The typical experimental conditions
were the following: $n_0 \approx 70$, $\epsilon_0 
= \epsilon n_0^{4} \approx 0.05$, $\omega_0 = \omega n_0^{3} \approx 1$ 
where $n_0$ is the principal quantum number of initially excited state,
$\epsilon$ and $\omega$ are the strength and the frequency of microwave field
(here and below we use atomic units). The classical dynamics depends only
on the rescaled values $\epsilon_0$ and $\omega_0$.

\indent
For the understanding of the process of ionization 
in linearly polarized field it is convenient to use
the one-dimensional atom model \cite{JEN,4M,EEE}. The investigations
of one-dimensional model showed that for high microwave frequency
($\omega n^3 > 1$) 
the dynamics of the system, which 
originally is ruled by the continuous Hamiltonian equations, can be
described by the Kepler map \cite{EEE}:
%1
\begin{equation}
\bar{N}=N+k\sin {\phi},\ \   
\bar{\phi}=\phi+2\pi\omega(-2\omega\bar{N})^{-3/2}
\end{equation}
Here $k=2.58\epsilon/\omega^{5/3}$,  $N=E/\omega$ 
has the meaning of the number
of absorbed or emitted photons ($E$ is the energy of the electron), $\phi$ is
the phase of microwave field at the moment when 
the electron passes near the nucleus. The bar denotes the new values of
the variables after one orbital period. 
\par
The physical reason due to which the motion 
can be quite accurately \cite{EEE} described by the 
simple area-preserving map
is the following: when the electron is far from the nucleus microwave 
field leads only to a small fast oscillations which doesn't modify the
average energy and the Coulomb trajectory of the electron. The change of
energy happens only at perihelion where the Coulomb singularity leads to 
a sharp increase of the electron velocity. Ionization takes place 
when the energy of the electron becomes positive after a pass near the
nucleus $N>0$. Then the electron goes to infinity and never returns back.
Therefore for the map (1) 
ionization is equivalent to absorption of trajectories with
$N>0$.
\par
To find the chaos border in the Kepler map we can linearize the second 
equation in (1) near the resonant (integer) values of $\omega n^3$
obtaining the Chirikov standard map \cite{10M}:
%2
\begin{equation}
\bar{N}=N+k\sin {\phi},\ \   
\bar{\phi}=\phi+T\bar{N}
\end{equation}
with $T=6\pi{\omega^2}{n}^5$. The global chaos appears for $K=kT>1$
that determines the critical field strength above which the classical 
atom is ionized. In this regime excitation goes in a diffusive way with the
diffusion rate $D=(\Delta N)^2/{\Delta \tau}=k^2/2$ where $\tau$ measures
the number of orbital periods of the electron.
\par
The first numerical and analytical investigations of the quantum 
one-dimensional atom model \cite{4M} showed that quantum effects leads to
the suppression of classical diffusion. Indeed, 
in the quantum case the variables $(N,\phi)$ becomes the operators with the
commutation rule $[N,\phi]=-i$ and the system is locally described by the 
quantum kicked rotator \cite{11M}. The photon number is analogous to the
level number in the kicked rotator and
the excitation probability decreases
exponentially with the number of absorbed photons 
so that the ionization
rate is proportional to $W_{I} \sim \exp(-2 N_{I}/l_{\phi})$. Here
$N_{I} = n_0/2{\omega_{0}}$ is the number of photons required for ionization,
$l_{\phi} = D =3.33\epsilon^2/\omega^{10/3}$ is the localization length.
For $l_{\phi} << N_{I}$ quantum ionization is exponentially small in 
comparison
with the classical value. However, for  $l_{\phi} > N_{I}$ the
diffusion is delocalized and the process of ionization is close to the
classical one. 
\par
In the 3-dimensional atom the Coulomb degeneracy leads to a slow
motion along energy surface that allows to describe the excitation in energy
also by the Kepler map with a small change of constant $k$. The motion
along the energy surface has some additional integral of motion that
explains the existence of localization in 3-dimensional atom
\cite{EEE,25D}. Recently the existence of
localization in the 3d case was reconfirmed in \cite{ANDR}.  
\par
Quantum localization of classical chaotic ionization has been observed in the
microwave experiments with hydrogen \cite{13M,14M} and rubidium
\cite{15M} atoms. Numerical simulations with the quantum Kepler map
\cite{12M} reproduce the  $10\%$ ionization threshold obtained in the
laboratory \cite{13M}.  The theoretical prediction for the quantum
delocalization border was also observed in the skilful numerical simulations
\cite{ANDR}. 
\par
Being very successful in the description of energy excitation
the Kepler map, however,  cannot be applied for the case of very strong
field. Indeed, in its derivation it was assumed that the change of energy
after one kick $k \omega$ is much larger than the energy of free oscillations
$\epsilon^2/2\omega^2$. This gives the condition of applicability
of the Kepler map picture \cite{EEE}: 
%3
\begin{equation}  
\epsilon<<\epsilon_{ATI} \approx 5\omega^{4/3}
\end{equation}
Let us note that this condition is independent on the initial state
since $n_0$ doesn't enter directly in the expression for $\epsilon_{ATI}$.
\par
In the one-dimensional case for $\epsilon>>\epsilon_{ATI}$ a collision with
the nucleus, being unavoidable, goes in a fast way like with an elastic 
wall leading to a prompt ionization \cite{EEE}. In the two-dimensional case
for zero magnetic quantum number $m$ such collision also always takes place
if the amplitude of free oscillations $\epsilon/2\omega^2$ 
is larger than the unperturbed distance between the electron and the nucleus
in perihelion ${l}^2/2$ ($l$ is the orbital momentum). This gives
the condition of prompt ionization for $l>(3/\omega)^{1/3}$ \cite{8M}:
%4
\begin{equation}  
\epsilon>\omega^2 l^2/4
\end{equation}   
where it was assumed that $l$ is few times less than $n$.
For $l<(3/\omega)^{1/3}$ ionization is ruled by the Kepler map and for
$\epsilon_{0} > {{\omega_{0}}^{2/3}}/2.6$ prompt ionization takes place after
one orbital period (see (1)).
Therefore, there is no stabilization of classical atom in the strong field 
for $m=0$. However, for high orbital momentum the atom remains stable
up to very high field values.  

\section{Stabilization Border}
\indent
\par
While for the magnetic number $m=0$ ionization always takes place
in a sufficiently strong field the case of nonzero $m$ is much more
interesting. Indeed, for the linear polarization of the field the 
projection $m$ is an exact integral of motion and the created by it 
centrifugal potential gives a possibility to avoid a collision with the 
nucleus.
To analyze the motion in the strong field
it is convenient to use the oscillating Kramers - Henneberger 
frame \cite{ALL} and cylindrical coordinates
in which the Hamiltonian has the form:    
%5
\begin{equation}  
{H} ={{p_z}^2\over 2}+{{p_{\rho}}^2\over 2} 
 + {{m}^2\over {2{\rho}^2}} - 
{1\over {({\rho^2}+
{({z}-{{\epsilon}\over {\omega}^2}\sin(\omega t))}^2)^{1/2}}}
\end{equation}   
If the frequency of the nuclear oscillations is large enough
(the condition will be given later) then in first approximation
the nucleus can be considered as a charged thread with a linear 
charge density $\sigma$ slowly dependent on $z$: $\sigma (z)=
\omega^2/(\pi \epsilon (1-(z \omega^2/\epsilon)^2)^{1/2})$. 
Then, for small $z$ and $\rho$ the Hamiltonian of averaged motion 
takes the form \cite{DS,BCS}:
%6
\begin{equation}  
{H_{ave}} ={{p_z}^2\over 2}+{{p_{\rho}}^2\over 2} 
 + {{m}^2\over {2{\rho}^2}} 
 +{{2 \sigma (z)} \ln ({{\rho \omega^2} \over {\epsilon}})} 
\end{equation}   
The constant under the logarithm takes into account that for 
$\rho >> \epsilon/\omega^2$ the coupling energy becomes much
less than $\omega^2/\epsilon$. From (6) one easily finds the
position of the potential minimum 
$\bar{\rho}=\sqrt {\pi \epsilon/2} m/\omega$ and 
the frequency of small oscillations 
$\Omega=2 {\sqrt 2} \omega^2 /(\pi \epsilon m)$ (for 
$z<<\epsilon/\omega^2$). The depth of the potential or the 
energy required for ionization of atom is approximately
$I \approx {{2 \omega^2} L/ {\pi \epsilon}} $ with
$L= {\ln (2 \epsilon /(e \pi \omega^2 m^2))/2}$. 
The minimal distance between the nucleus and electron is 
determined by the condition $I= m^2/2 {(\rho_{min})}^2$ giving:
%7
\begin{equation}  
\rho_{min}={m \over 2\omega} {\sqrt {\pi \epsilon \over L}}
\end{equation}   
The physical reason for the growth of the minimal distance with
the field strength is the following: with the increase of the field
the amplitude of the field oscillations grows leading to the
decrease of attractive Coulomb force while the centrifugal 
potential remains the same. 
\par
The averaged description of the motion (6) is correct if the
frequency of field oscillations $\omega$ is much larger than 
the frequency $\Omega$ of oscillations in $\rho$. In that case the 
averaged Hamiltonian (6) is the constant of the motion with adiabatic 
accuracy and ionization of atom doesn't take place. This gives the
stabilization border \cite{DS,BCS}:
%8
\begin{equation}  
\epsilon > \epsilon_{stab}=\beta {\omega \over m}
\end{equation}   
where $\beta$ is some numerical constant. The same estimate can be obtained
from the condition that the change of energy $\Delta E$ 
during the collision between the electron and the nucleus is smaller 
than $I$. Indeed, the change of the momentum is 
${\Delta p} \approx {\Delta t/{\rho_{min}}^2} 
\approx {\omega /(\epsilon \rho_{min})}$ and the change of the energy
$\Delta E \approx {\omega^2}/(\epsilon^2 {\rho_{min}}^2)$ is
less than $I$ if (8) is satisfied. It is interesting to note that the
stabilization border (8) can be written as 
$v_{n}=\epsilon/\omega >  v_{max}$ where $v_{n}$ is the typical velocity
of the nucleus and $v_{max}=2/m$ is the maximal velocity of the electron
in the atom without the external field. 
\par
Another condition intrinsically used 
in the derivation of (6) and (8) is  $\rho_{min}<\epsilon /\omega^2$
which gives $\epsilon>m^2 \omega^2$. Also, there are two qualitatively
different situations depending on the ratio between $m$ and 
$(3/\omega)^{1/3}$. In the case $m<<(3/\omega)^{1/3}$ 
(stabilized atom regime) we have 
$m^2 \omega^2 << 5 \omega^{4/3} << \beta \omega/m = \epsilon_{stab}$.
For small field amplitude (3) the excitation is described by 
the Kepler map and the complete ionization after one orbital period of the
electron takes place for $\epsilon_{0}>{\omega_{0}}^{2/3}/2.6$
\cite{EEE}. Between this border
and above chaos border $\epsilon_{c0} = 1/49 {\omega_{0}}^{1/3}$
ionization goes  in the diffusive way which is also relatively fast.
However, for the more strong field (8), when the Kepler map
picture is not valid (see (3)), atom becomes stable. 
The case of opposite
inequality is less impressive. Indeed, for $m>>(3/\omega)^{1/3}$ 
(stable atom regime) we
have $\beta \omega/m<< 5\omega^{4/3}<<m^2 \omega^2$ and atom remains
stable (nonionized) up to $\epsilon \sim {m^2 \omega^2}$ as it was
in (4) ($l \sim m$). Above this value a significant portion 
(order of half) of atoms will remain stable since condition (8) is
satisfied. Finally, ionization takes place only when the value of
$\rho_{min}$ (7) becomes larger than the size of the atom $2 {n_{0}}^2$
and the electron cannot be captured in the stable region during the
switching of the field. This gives the destabilization border
%9
\begin{equation}  
 \epsilon_{destab} \approx {{16 L \omega^2 {n_{0}}^2 }\over {\pi m^2}}
\end{equation} 
This border is also valid for the case $m<<(3/\omega)^{1/3}$. Of course,
in that case the stabilization can be observed only for the time of
field switching $T_{sw}$ less or order of one orbital period 
of the electron. Otherwise a collision with nucleus will take place
at field strength $\epsilon < \epsilon_{stab}$ and atom will be ionized.
\par
The results of numerical simulation of ionization process of system
(5) are presented on the Fig.1. The stabilization
probability $W_{stab}=1.-W_{ion}$  
is given for different field strengths 
$\epsilon_{0}$ and frequencies $\omega_{0}$.
The ionization probability $W_{ion}$ was defined as the relative part 
of the trajectories with positive energies after field pulse.
The initial distribution of 100 trajectories (25 for $\omega_{0}=1000.$)
corresponded to a quantum state with fixed spherical quantum numbers
(fixed actions and equipartition in conjugated phases). The initial 
value of orbital momentum was equal to $l/n_{0}$=0.3 and its projection 
was equal to $m/n_{0}$=0.25.  The time of field switching (on/off) 
measured in the number of field periods was chosen to be 
equal to $T_{sw}=\omega_{0}$ 
(one unperturbed orbital period of the electron). The pulse duration
of the field was $T_{int}=500 \omega_{0}$ (500 orbital periods). 
The data clearly demonstrate the stabilization of atom for field
strength larger than some critical value. It is convenient to define 
the stabilization border as the field strength 
$\epsilon_{stab0}(20\%)$ for which $W_{stab}=0.2$. The dependence of
$\epsilon_{stab0}(20\%)$ on $\omega_{0}$, extracted from the data of Fig.1
can be well fitted by the theoretical expression (8) with $\beta=12$ 
in the wide frequency range (see Fig.2 of \cite{DS}). 
This dependence continues up to 
$\omega_{0}=1000$ where we enter in the stable atom regime with
$m>(3/\omega)^{1/3}$ and where stabilization disappears in agreement with
above theoretical arguments (see Fig. 1). However, the stability of
atom in that case is of the other nature than it was in \cite{8M}
since the condition (4) is strongly violated. So, for such strong fields
the stability of atom is based on the same physical grounds (8) as in the
stabilized atom regime for $m<<(3/\omega)^{1/3}$.
The numerical check of the dependence of stabilization border 
$\epsilon_{stab0}(20\%)$ on $m$ as well as the destabilization
border (9) on $\omega_{0}$ 
demonstrates good agreement with the theory (8)-(9) \cite{DS}.

\section{Kramers Map}
\indent
\par
It is important to stress that according to (8) the stabilization 
can take place even when the size of electron oscillations 
$\alpha=\epsilon/{\omega^{2}}$ is much less
than the unperturbed size of the atom $n_{0}^{2}$. An example of the
motion in this case is presented on the Fig.2.
In such a case the electron follows the usual Kepler elliptic orbit
and his energy (the size of the orbit) can be changed only during
his fast passage near the nucleus. In this sense we can expect that 
the motion can be effectively described by some map analogous to the
Kepler map.
\par
To construct such kind of map let's introduce a simplified 
one-dimensional Kramers model given by the Hamiltonian
%10
\begin{equation}  
{H} ={{p_{\rho}}^2\over 2} 
 + {{m}^2\over {2{\rho}^2}} - 
{1\over {({\rho^2}+
{{{\epsilon}^{2}\over {\omega}^4}(\sin\gamma+\sin(\omega t))}^2)^{1/2}}}
\end{equation}   
This model is obtained from the Hamiltonian (5) by neglecting the
changes of $z$ and considering $z=-\epsilon/\omega^{2} \sin\gamma$
as a constant. In other words electron collides with the line $\rho=0$
always at the same $z$ value. 
The physical reason for that is the following.
The collision of the electron with the nucleus is analogous to
a collision of a fast heavy particle with the light electron.
In such collision the change of energy (velocity) takes place mainly
in the perpendicular $\rho$-direction, while the velocity
in $z$-direction remains practically the same. 
According to this physical picture the model (10) mainly presents
the changes in $\rho$-direction. In that sense it is quite different
from the well known one-dimensional atom model of Eberly \cite{ALL}
which implicitly takes into account the change of energy (velocity)
only in $z$-direction. Also in \cite{BALA} the authors considered
the velocity change only in $z$ that has led them to a higher
stabilization border than (8), while the estimate for $\rho_{min}$
has been found correctly (see (7)).  
\par
According to the analogy with the Kepler map we can expect that the
change of the electron energy in the model (10) will take place only
when the electron passes near the nuclear and that it will depend only
on the phase of the field $\phi = \omega t$ at that moment. 
If also the size of the
orbit is much lager than the size of the nucleus oscillations
($\alpha=\epsilon/\omega^{2} << {n_{0}}^2$) then the change of
the phase is given by the Kepler law and is the same as in (1).
Basing on these arguments we can assume that the dynamics of energy 
excitation is governed by the Kramers map of the following form:  
%11
\begin{equation}
\bar{E}=E+J h({\phi}),\ \   
\bar{\phi}=\phi+2\pi\omega(-2\bar{E})^{-3/2}
\end{equation}
with $E=\omega N$, where as in (10) $N$ is the photon number,
the maximum change of the energy is given by a constant $J$ and the 
unknown function of the kick $h({\phi})$ varies in the interval [-1,1]. 
\par
To check the validity of this map I integrated the continues equations of
motion for the model (10) and plotted the change of energy as a function
of the field phase at the moment when the value of $\rho$ took
one of its minimal values ($p_{\rho}=0$). Such approach allows to
find the kick function $h(\phi)$ the examples of which are presented
on Figs.3,4. The numerical results clearly demonstrate that the
function $h$ exists. However, it has a quite unusual property.
Indeed, some values of $\phi$ never appear (even if the number of periods
was increased in 20 times). These values of $\phi$ are approximately
equal to $\pi+\gamma, 2\pi-\gamma$ and correspond to that values of 
the field at which the nucleus passes via the point of collision
$z=-\epsilon/\omega^2 \sin\gamma$. A more close consideration of motion
near these special $\phi$-values shows that the electron remains during
some small time interval (within corresponding phase interval $\Delta \phi$) 
near the nucleus making one (Fig.3) or
two (Fig.4) oscillations in $\rho$ of very small amplitude so that
the value of $\rho$ remains practically (but not exactly) the same.
This gives correspondently two (or three) values of the phase $\phi$
with the same change of $\Delta E$ since the value of $E$ was determined
in the aphelion. This of course puts the question about the derivation
of the Kramers map in some other synonymous form. However, the main properties
of the motion can be derived already from the approximate representation  
(11) where we will define the function $h$ in the empty intervals by
connecting the last points at the ends of the interval by straight line.
\par
Defined in such a way the Kramers map has the properties quite similar
to the Kepler map. Indeed, the function $h$ is close to $\cos\phi$
and the approximate chaos border in (11) can be defined by the linearization
of the second equation giving:
%12
\begin{equation}  
K=6 \pi \omega J n^5 > 1
\end{equation}   
where we used substitution $E=-1/2n^{2}$. According to this criterion
and in agreement with the numerical data the motion is chaotic
for the cases of Figs. 3,4. If to introduce $k=J/\omega$,
which will give the number of absorbed photons after an orbital period,
we will get the same formulas for the diffusion rate $D=k^2/2$,
the localization length ($l=D$) and the ionization time $\tau_{ion}=
N_{I}^2/D$ as in the Kepler map. In this sense the most important
problem is the definition of the dependence of $J$ on the parameters
of the system.
\par
According to the results of the previous section the amplitude of the kick
$J$ must decrease exponentially  with the increase of the 
stabilization (adiabatic) parameter
$S=\epsilon m/\omega \sim \omega/\Omega$ (see (7),(8)). This expectation
is in agreement with the results presented on Fig.5. Indeed, the exponential
decrease of $J$ with the field strength, and therefore stabilization, 
are evident. Let us at first discuss the properties of $J$ for
nonzero values of $\gamma$. Even though the value of energy for the cases
of Fig.5 
was quite small nevertheless there is some dependence of $J$ on energy.
An example of such dependence is presented on the Fig. 6. We see that 
$lnJ$ depends on energy $E$ approximately in a linear way and goes
to a constant value for $E=0$. This means that in the limit
$n_{o}^2>>\epsilon/\omega^2$ the value of $J$ is independent from $n_{0}$. 
This result is consistent with the above arguments that the change of the
energy takes place only in the small vicinity of the nucleus. However,
in the difference from the Kepler map it is necessary to have quite
strong inequality $-\alpha E << 1$ to neglect the dependence of $J$ on $E$.
We will try to explain this fact later. In this regime of small energies
the main change of the phase of the field between collisions (the second
equation in (11)) is obviously given by the Kepler law.
\par
To determine the dependence of $J(E=0)$ on the parameters it is convenient to
fix the stability parameter $S$ that allows to eliminate the strong
exponential dependence and to find the factor before the exponent.
The numerical results are presented on Fig.7.
The values of $J(E=0)$ were obtained from nonzero energies by linear
extrapolation to $E=0$ (see Fig.6). The numerical data clearly show
that for the fixed $S$ the value of $J(E=0)$ is independent on the
frequency and is inversely proportional to $m^2$. In principle
the factor $1/m^2$ gives simply the correct dimensionality however
the independence on the other dimensionless parameter $\nu=m\omega^{1/3}$
is not so obvious.
\par
Combining all the obtained numerical results we can present the dependence
of the kick amplitude $J$ on the parameters for 
$\mid E \mid \epsilon/\omega^{2} <<1$ in the following form: 
%13
\begin{equation}  
J={g_{1} sin\gamma \over {m^2}} 
\exp({-  (g_{2} - g_{3}\epsilon E/\omega^{2}})\epsilon m/\omega)
\end{equation}   
where $g_{1,2,3}$ are some functions weakly dependent on $\gamma$. 
For $\gamma=0.6$ we
have from Figs. 5-7 that $g_{1} \approx 0.13$, 
$g_{2} \approx 0.19$, $g_{3} \approx 0.08$.  The numerical data for other 
values of $\gamma$ show that the fitting parameters vary not more than in 2 
times for practically the whole interval of $\gamma$.
For example, $g_{1}$ = 0.1 and 0.2, $g_{2}$ = 0.13 and 0.21,
$g_{3}$ = 0.045 and 0.1 for $\gamma$ = 1.2 and 0.3 correspondingly.
\par
To understand the numerically obtained formula (13) for $J$ it is possible to
make the following estimate. Taking the partial time derivative from the
Hamiltonian (10) we obtain the expression for the change of the energy
after one orbital period:
%(14)
\begin{equation}  
\Delta E =  {\epsilon^2 \over \omega^{4}}
\int {{\cos(\eta+\phi) ({\sin\gamma+\sin(\eta+\phi)})} d \eta  \over 
{\rho^{3/2}(\eta)}}
\end{equation}   
where $\eta=\omega t$ and in the denominator we neglected the term with 
$\epsilon^2/\omega^4$ in comparison with $\rho^2$.
We can assume that near the nucleus the time dependence
of $\rho$ is the same as for a free electron with momentum $m$
that gives $\rho^2(t) = \rho_{0}^{2}+v^{2} t^{2}$ where $\rho_{0}$
is the minimal distance from the center and $v$ is the velocity of the 
electron far from the center. For this free motion with the fixed momentum
$m$ we have the relation: $\rho_{0}^{2} = m^2/v^2$. For the velocity
it is possible to use the following expression: $v^2={\omega^2/{C \epsilon}
+2 E}$. Where the first term takes into account the fact that the energy must 
be
measured in respect to the minimum of the effective potential (see (6)) and
$C$ is some unknown constant. It is easy to see that $C$ determines the 
minimal
distance $\rho_{0}^{2} = C \epsilon m^2 / \omega^{2}$ for $E=0$. In principal 
the value of $C$ depends on $\gamma$.
\par
After substitution of all these expressions in (14) we obtain the following
estimate
%15
\begin{equation}  
J \sim {{S^2} \over {m^{2} (m^{3/2} \omega^{1/2})}} 
\sin\gamma exp(-C S / (1+2 C \epsilon E/\omega^2)), \ \
h(\phi)=cos(\phi), \ \ S=\epsilon m / \omega  
\end{equation}   
Of course, the presented derivation is not exact. However, it
reproduce quite well the exponential dependence (13)
(while the factor before the exponent is not in agreement with
the dependence obtained from the numerical simulation). 
Indeed, numerically $h(\phi)$
has maxima near $0$ and $\pi$. Comparison of (15) with (13) gives
$g_{2}=C$ and $g_{3}=2 C^2$. The value of $C$ can be defined directly 
from the numerical simulation of the one-dimensional Kramers model
for different $\gamma$. The comparison of the $g_{2}$ with $C$ is presented on
the Fig.8 showing the good agreement with the prediction.
The ratios of the numerical value of $g_{3}$ (see above) to the theoretical
value $2 g_{2}^{2}$ are equal to 1.13, 1.1, 1.33 correspondingly for 
$\gamma=0.3, 0.6, 1.2$ and are also in good agreement with the theoretical
estimate. In the future estimates we will use the expression
(13) with the theoretical substitution for $g_2$ and $g_3$.
Let us also mention that for $\gamma=0$ we get from (14) that 
$g_2 = 2 C$ (from Fig.5 the ratio to the theoretical
value is approximately 1.1) and $h(\phi)=\sin(2\phi)$ that is quite
close to the numerical data. Further theoretical analysis is required to
obtain the factor before the exponent in (13).
\par
While still there are some unclear questions with the construction of the
Kramers map the approximate consideration made above and the analogy with
the Kepler map allow to understand the main properties of motion.
If the number of photons required for the ionization is large then, as it was
for the quantum Kepler map, it is possible to have diffusive excitation
and quantum localization of chaos. However, due to high values of the 
frequency it is also quite easy to have a situation when one photon
can already lead to ionization. In this case 
for $k=J/\omega<1$ the one-photon ionization rate (per unit of time) 
is given by the perturbation theory and
as for the Kepler map (see \cite{EEE}) it is equal:
%16
\begin{equation}  
\Gamma \approx {J^2 \over {8\pi n^3 \omega^2}}
\end{equation}   
For $J>w$ approximately a half of probability is ionized after one orbital 
period ( in (11) as in the Kepler map the orbit is ionized if
after a kick $E>0$). From (16) it is clear that we may have long living states
if the field is sufficiently strong. From the quantum view point one of the
most interesting cases is the case of small $m$. In this case we need to
make substitution $m \rightarrow m+1$ since as it is well known the
correct quasiclassical quantization leads to the appearance of the effective
centrifugal potential even for zero orbital momentum. That gives the
stabilization border $\epsilon > 10 \omega $ for $m=0$.
\par
Finally let us mention that in (11) we assumed that $J$ is independent on
energy. To take into account this dependence we need to put in the first
equation $J=J(\bar{E})$ and in the second equation to add the phase shift
$\Delta \phi = d J/d{\bar{E}} f(\phi)$ with $h(\phi)=-df(\phi)/d \phi$.
In that way the map will remain canonical.

\section{ Channeling Analogy }
\indent
\par
Here I would like to discuss the analogy between the phenomenon of
stabilization of atom in strong field and the channeling of particles in
a crystal (see for example \cite{16M} and Refs. there in). 
Let's consider the electron moving in the crystal with the velocity
$v \approx c=137$ (we will consider nonrelativistic case). Then in the 
frame of the moving electron its interaction with the protons in the crystal
lattice will have approximately the form (5) if to take into account the
interaction only with a nearest proton. 
On the grounds of that analogy we find
that the effective distance between atoms in the crystal $a$ 
and the velocity of the electron are equal to:
%17
\begin{equation}  
a={\epsilon \over \omega^2}, \; v={\epsilon \over \omega}
\end{equation} 
The frequency of perturbation is $\omega=v/a$ so that $\epsilon=v^2/a$. 
Since in the crystal the distance between the atoms is approximately the same
in all directions 
the analogy is valid for $\epsilon / \omega^2 > {n_0}^2$. 
The necessary condition of channeling is that the critical injection angle
$\theta$ must be much less than one that implies: 
$\theta \approx v_{\perp}/v \approx 1/(v \sqrt{a} )
\approx (\omega^{4/3}/\epsilon)^{3/2} << 1$. This is the condition of
unapplicability of the Kepler map (3). From the stabilization condition
(8) it follows that channeling takes place for electrons with momentum
$m>10/v $. This is always satisfied for fast electrons with $v \approx 137$. 
The existence of channeling for very energetic electrons
(that corresponds to strong field for stabilization problem) gives one more
evidence for existence of stabilization of atom in strong field in the regime
when one photon frequency is larger then the ionization energy. 

\section{ Conclusion }
\indent
\par
Basing on the Kramers map (11) and using its analogy with the Kepler map
we obtained the estimate for the one-photon ionization rate (16). This
ionization rate sharply  decreases with the stabilization parameter
$S=\epsilon m/\omega$. Such stabilization for excited states has some
interesting advantages in comparison with the stabilization of atom in the
ground state. Indeed, in this case stabilization can take place with
$\epsilon << 1$ and $\omega << 1$. This leads to a large energy difference
$\delta E$ between the excited states and the ground state. So,  the energy 
in an exited state is approximately $(\epsilon/\omega)^2/2>>1$ while
the energy of the ground state remains as in the unperturbed atom
(it is not the case for $\epsilon$, $\omega >>1$ when the ground state is 
also stabilized since there the electron has the same energy of free 
oscillations).
Due to that in the case of Rydberg stabilization it is possible to have
radiative transitions to the ground state with the radiation of X-ray
photons. For the frequency of $CO_{2}$ laser $\omega \approx 1/300$ (0.1 ev)
and $m$=0 (or 1) the stabilization will take place for $\epsilon \approx 1/30$
(1.6 $10^8$ V/cm). The size of the atom will be larger than the size of
the field oscillations $\alpha=\epsilon/\omega^2$ for $n>40$. According
to (13) and (16) for the field $\epsilon$ = 5 $10^8$ V/cm and $n$=60
the life time of the atom will be about 5 $10^5$ orbital periods
or $10^{-6}$ seconds (we take for the estimate the case with $\gamma=0.6$).
Of course, to obtain such states the time of field switching must be 
less than the time of orbital period
as we discussed above. Since recently it was predicted that
the Rydberg atoms can form long living states (bands) in the solid state
\cite{JETP} (giving very high density of excited atoms) it will be interesting
to consider a possibility of stabilization not only for a separate atom but
also for such Rydberg solid state.
\par
I had started to be interested in the problem of microwave ionization
of hydrogen atom in far cold 1980 during the first visit of Jeff Tennyson to
the group of Boris Chirikov at Novosibirsk. Now, after many years of 
researches by different groups, this problem still 
continues to live by its own life as well as the memory about Jeff
continues to live among the people who met him in Siberia.

\vfill\eject

\renewcommand{\baselinestretch} {2}

\vfill\eject
{\bf {Figure captions}}
\vskip 20pt
\begin{description}
{
\item[Fig. 1:] Stabilization
probability $W_{stab}=1.-W_{ion}$  
is given for different field strengths 
$\epsilon_{0}$ and frequencies ($\omega_{0}$=0.3 ($\circ$), 
1. ($\ast$), 3. (+), 10. ($\Diamond$), 30. ($\triangle$), 
100. ($\Diamond$), 300. ($\triangle$), 1000 ($\bullet$)). 

\item[Fig. 2:]  Example of trajectory  for $\omega_{0}=300$, 
$\epsilon_{0}=20000$ with initial $l/n_{0}=0.3$ and $m/n_{0}=0.25$; 
 $10^{5}$ field periods are shown.

\item[Fig. 3:] Example of numerically obtained kick function $h(\phi)$
in Kramers map (11) for $\epsilon = 3$ $10^4$, $\omega$ = 125, $m$=0.2, 
$\gamma$=0.6, $E$=-0.125 (so that effective $n_0$=2), $J$=1.1 $10^{-4}$. 
Near 200 orbital periods (points) are shown.

\item[Fig. 4:] The same as Fig.3 with 
$\epsilon = 4 10^4$, $\gamma$=1.2 and $J=5.8 10^{-4}$.

\item[Fig. 5:] Dependence of the kick amplitude $J$ in (11) on stabilization
parameter $S=\epsilon m / \omega$ for $\omega=125$, $m=0.2$, $\gamma=0$ 
($\circ$); $\omega=1000$, $m=0.1$, $\gamma=0$ (+); 
$\omega=125$, $m=0.2$, $\gamma=0.3$ (open squares);
$\omega=125$, $m=0.2$, $\gamma=0.6$ (points);
$\omega=125$, $m=0.2$, $\gamma=1.2$ (full squares). For all cases $E$=-0.125.
Lines are drawn to adopt an eye.  

\item[Fig. 6:] Example of dependence of $J$ on energy $\mid E \mid$ for
$\epsilon = 2.2$ $10^4$, $\omega=125$, $m=0.2$, $\gamma=0.6$ (points).

\item[Fig. 7:] Dependence of $J$ on $m$ for fixed stabilization parameter
$S=35.2$ and $E$=0; 9 cases are shown for $\omega$ in the interval
[10,1000] and $m$ in the interval [0.05,0.6]. The straight line
shows the dependence $J \sim 1/m^2$.

\item[Fig. 8:] Dependence of $C=\rho_0 \epsilon/S^2$ on $\gamma$
(full line). Points give values of $g_2$ to demonstrate theoretical
relation $C=g_2$.

}
\end{description}
\vfill\eject
\end{document}